\documentclass[%
reprint,
 amsmath,amssymb,
 aps,
]{revtex4-2}

\usepackage{graphicx}
\usepackage{xcolor}
\usepackage{dcolumn}
\usepackage{bm}

\begin{document}

\preprint{APS/123-QED}

\title{On repeated measurements of a quantum particle in a harmonic potential}

\author{Filip Gampel}
\author{Mariusz Gajda}%
\affiliation{%
 Institute of Physics Polish Academy of Sciences\\
 Aleja Lotnikow 32/46, 02-668 Warszawa, Poland 
}%




\date{\today}

\begin{abstract}We study evolution of a quantum particle in a harmonic potential whose position and momentum are repeatedly monitored. A  back-action of measuring devices  is accounted for. Our model utilizes  a generalized measurement corresponding to the Positive Operator-Valued Measure. We assume that upon measurement the particle's wavefunction is projected onto one of possible detector states depending on the observed result. We chose these post-measurement states to be moving Gaussian wavepackets. The  Wave Function Quantum Monte-Carlo formalism is used to simulate single quantum trajectories of the particle. We show how classical trajectories emerge in course of observation and study in detail   dispersion of position and momentum of the particle.  


\end{abstract}

\maketitle


\section{\label{sec:1}Introduction}
 
Position and momentum are fundamental quantities characterizing the dynamics of a classical particle. The time-dependent position of a particle is directly related to what an observer sees while monitoring its motion. The concept is thus very intuitive. According to classical mechanics, measurements in principle do not affect the system, and their precision can be arbitrarily high. In contrast, quantum mechanical measurements always somehow affect the system, and moreover, the relation between a wavefunction (or density operator) describing the state of a system to what is actually being observed is not so obvious. 

The first approach to resolve these issues  is known as the Copenhagen interpretation   \cite{Born1927,Heisenberg1930,vonNeumann1932,Peres}, which, until today, forms the basis for the textbook version of quantum mechanics. A central role is played by the Born rule, which gives probabilities of  positive answers to yes/no questions related to measurement outcomes. When a measurement is completed, an answer is obtained and the wavefunction changes discontinuously, in accordance with the result and the von Neumann (and L\"uders) postulate of wavepacket reduction \cite{vonNeumann1932,Luders51}.  P. Langevin expressed this rule in the introduction to the textbook `La theorie de l'observation en mecanique quantique' \cite{London39}) by F.~London and E.~Bauer, in the following words:  {\it  `The wave function it [the quantum theory]  uses to describe the object no longer depends solely on the object, as was the case in the  classical representation, but, above all, states what the observer knows and what, in consequence, are his possibilities for predictions about the evolution of the object. For a given object, this function, consequently, is modified in accordance to the information possessed by the observer'}. 
\\

The Copenhagen interpretation gives a well defined prescription on how to use the theory in practice. However, it is not the only existing interpretation of concepts such as wavefunction and measurements. After decades, the issue of collapse, nonlocality and measurement still remains a subject of scientific discussion \cite{Wigner63,Kraus83,Ludwig83,Holevo01,Zurek2003,Wiseman10}.\\

Iwo Białynicki-Birula  and Zofia Białynicka-Birula (Z-IBB) identify in their  textbook `Quantum electrodynamics'\cite{IBB} the fundamental postulates of quantum theory pertaining to the relation between the density operator and measurement. The postulates are very formally combined into four axioms which can be, under some simplifications, summarized as follows: i) the elementary questions, i.e. the yes/no questions, are represented by projectors, $P$; ii) the state of a system is represented by a non-negative, self-adjoint, and trace-one density operator $\rho$; iii) the density operator determines probabilities $p$ of affirmative answers to elementary questions in accordance with the Born rule $p=Tr\{P\rho\}$; iv) every dynamical variable $\cal A$ is represented by a self-adjoint operator $A$, and can be assigned a spectral family of projectors, $E^{(A)}_\lambda$ symbolizing questions whether the value of a dynamical variable $\cal A$ is not larger than $\lambda$. 
As mentioned by Z-IBB, {\it `since the set of probabilities $p$ is the only information in quantum theory available about the state of the  system, from the operational point of view the concept of state of the system should be identified with the function $p({\cal P})$ defined on the set of all questions.'} \\

The collapse postulate is missing from the Z-IBB axioms. One might possibly find it in the statement quoted above, equating the state of the system to the function $p({\cal P})$. The answer to any question apparently modifies it. On the other hand, the issue might have seemed purely academic at the time, since realistic measurements in quantum mechanics were generally believed to be destructive. This excludes the possibility of repeated measurement on the same quantum system and limits the relevance of the collapse postulate. E. Shr\"odinger \cite{Schrodinger}, one of the founding fathers of the quantum mechanics, wrote: {\it 'We never experiment with just one electron, or atom, or (small) molecule, we sometimes assume that we do; this inevitably entails ridiculous consequences $\hdots$.  In the first place, it is fair to say that we are not experimenting with single particles, any more than we can raise ichtyhysauria in the zoo'}. Nowadays, such measurements are not only theoretically considered, but also performed in labs. \\

 Modern variations of the Copenhagen interpretation, such as QBism -- Quantum Bayesianism \cite{Caves02,Fuchs10,Fuchs13,Mermin14}, postulate that an agent (e.g. a physicist) observing a system abruptly modifies their knowledge (the set of probabilities) once a measurement outcome becomes available. The wavefunction expresses the individual agent's state of knowledge.   {\it `...There is no real state of a physical system. What one chooses to regard as the physical system and what  state one chooses to assign to it depend on the judgment of the particular physicist who questions the system and who uses quantum mechanics to calculate the probabilities of the answers.'}, as stated by N. David Mermin \cite{Mermin2022}. According to Qbists, since there is no objective wavefunction of the system, there is no collapse either. \\

Other points of view assume that the wavefunction, the state of the system, has  attributes of reality, being independent of an observer. The issue of an apparent collapse, disliked by many physicists, is resolved in various ways. Everett's many-world interpretation (MWI) is one such approach \cite{Everett57}. Non-local hidden variable theories, Bohmian mechanics being the best known example, are another possibility \cite{Bohm}. The MWI postulates that upon measurement, the system, which finds itself entangled with the measuring apparatus, does not collapse to some observed state, but rather that all components of the wavefunction associated with possible measurement outcomes continue to evolve according to the Schrodinger equation of the composite system. Because of the linearity of the Schrodinger equation these components do not interact and form seperate `branches' or `worlds'. One must accept uncountable copies of themselves and the world living different lifes. Bohmian mechanics introduces additional hidden  variables, coordinates (e.g. particle positions) associated with a configuration of the system under consideration. The particles move guided by a `pilot wave', which is equivalent to the wavefunction of orthodox Quantum Mechanics (QM) and evolves according to the Schrodinger equation. It is thus the hidden variables which are actually observed in a measurement. Each of the varying interpretations of QM -- of which we only mentioned a few -- forces us to accept some non-intuitive, seemingly problematic postulate about reality. If none of them is found satisfactory, one must accept the view that collapse, an abrupt discontinuous change of the system, triggered by measurement, is a `real and wild' thing.\\

None of the interpretations presented above may be falsified on grounds of present knowledge. The problems, at this stage of understanding, seem of a philosophical nature, their experimental verification elusive. However, the various interpretations may imply effects measurable in future experiments and lead to different generalizations of quantum theory. \\

The first studies of repeated measurement of continuous variables can be found in works of Mensky \cite{Mensky79} and Davies, \cite{Davies69,Davies76}. Great experimental progress in cooling and trapping of single ions, opened many possibilities of repeated measurement of a  single quantum system. The first spectacular example is  observation of quantum jumps, i.e. dark periods in the fluorescence spectrum of an optically driven trapped ion \cite{Toschek78,Toschek86,Dehmelt86,Wineland86}. The experiments fueled the interest in the theory of repeated quantum measurements.  
The proper description of a system under repeated measurements calls for the inclusion of information gained, a back-action of the meters, as part of the dynamics. Different methods were developed  \cite{Caves86,Caves87,Barchielli82,Barchielli84,Gisin84,Diosi88a,Diosi88b}. \\

The  theoretical approach utilizes an open system formalism. It is based on the Gorini-Sudarshan-Kossakowski-Lindblad (GKSL) equation for the density operator \cite{Gorini76,Lindblad76}. This allows for studying all \emph{statistical} properties of the system. Instead of solving directly the GKSL equation, for different reasons it may be preferable to look for single realizations of a wavefunction dynamics. Obviously such individual trajectories are stochastic in nature. Averaged over many realizations, they provide a description equivalent to the time dependent density operator. The general theoretical framework governing wavefunction dynamics of this kind involves the introduction of the so-called stochastic Schr\"odinger equation (SSE) \cite{Gisin92,Carmichael93,Zoller95,Jacobs06,Ueda17}. It should be noted that the choice of an SSE is not unique and in general there are many realizations ('unravelings') corresponding to one GKSL equation. In fact, the formalism of SSE need not to be invoked at all for the construction of concrete numerical schemes generating the stochastical trajectories. One notable example \cite{Dalibard92,Molmer93} is known as the Wave Function Quantum Monte-Carlo (WFQMC) method. This formalism is often used by atomic physicists since it allows easily to generate sequences of events  mimicking experiments with atoms and photons.\\  

In this paper we use WFQMC to analyse statistical characteristics of trajectories determined by simultaneous repeated measurement of position and momentum of a quantum particle. First, we specify our model, we define jump operators and introduce the WFQMC approach. Then we present exemplary trajectories and discuss the time dependence of dispersion of position and momentum for different choices of detection parameters. Conclusions are presented in the final section.\\
 
\section{Monte Carlo dynamics of a wavefunction}
We study a phase-space trajectory of a quantum particle, continuously monitored by an array of detectors.  Here we use the theoretical model introduced by us in \cite{Gampel23}.  We assume that every measurement provides a value of position and momentum of the particle at this instant. A sequence of such readouts gives a phase-space trajectory. Each simultaneous measurement of position and momentum satisfies Heisenberg's uncertainty principle. We apply an open system formalism: our system is a quantum particle described by the Hamiltonian $H_0$, while the detectors form a reservoir. We assume that the reservoir has no memory. \\

The problem of simultaneous  measurement  of position and momentum was for the first time considered by E. Arthurs and J.L. Kelly \cite{Arthurs65}. The recent studies of A.J.  Scott and G.J. Milbourn \cite{Scott01} assumed a different detection model than studied here. They assumed the von Neumann type of coupling between a particle and a meter, and used a formalism based on an Ito stochastic Schr\"odinger equation \cite{Barchielli82,Barchielli84,Diosi88a,Diosi88b}. The main difference is, thus, in the form of the jump operators assumed here. \\

The effect of coupling of the system to the reservoir of detectors is described by the `jump operators' $C_{i,j}$ specified in the following part of the paper.  
The general form of a completely positive and trace preserving map which describes time-homogeneous dynamics of the density operator $\rho$ of a system coupled to the Markovian reservoir via operators $C_{i,j}$ is given by the Gorini-Kossakowski-Sudarshan-Lindblad equation: \cite{Gorini76, Lindblad76}:  
\begin{equation}
    \dot{\rho} = i [ \rho, H_0 ] + \mathcal{L}_{relax} (\rho),
    \label{master}
\end{equation}
where $H_0$ is the self-adjoint Hamiltonian of the system and $\mathcal{L}_{relax}$ is a relaxation operator of the Lindblad form, accounting for an effect of the environment:
\begin{equation}
    \mathcal{L}_{relax} (\rho) = - \frac{1}{2} \sum_\alpha \left(C_{i,j}^\dagger C_{i,j} \rho + \rho C_{i,j}^\dagger C_{i,j} \right) + \sum_\alpha C_{i,j} \rho C_{i,j}^\dagger.
\end{equation} 
We chose $C_{i,j}$ to be proportional to  projectors onto detector's states $|\alpha_{i,j}\rangle$: 
\begin{equation}
C_{i,j}= \sqrt{\gamma} |\alpha_{i,j}\rangle\langle \alpha_{i,j}|,
\end{equation}
where $\gamma$ gives the characteristic clicking rate (probability per unit time)
and $|\alpha_{i,j}\rangle$ are complex Gaussian wavepackets, which in position representations have the form:
\begin{equation}
\label{position}
    \langle x|\alpha_{i,j}\rangle = \frac{1}{(2 \pi \sigma^2)^\frac{1}{4}} e^{-\frac{(x-x_i)^2}{4 \sigma^2}} e^{i k_j x}.
\end{equation}
Spatial points $x_i$ and momenta $\hbar k_j$ define positions of the detectors in a phase-space. These locations are a matter of choice. Here we assume that they form a rectangular lattice with spacing $d_x$ and $d_p$ respectively. \\

In what follows  we will use the index $\alpha$ as  a shortcut notation for two indices, $\alpha \equiv (i,j)$ and $C_\alpha \equiv C_{i,j}$. The operators $C_\alpha$ are responsible for a reduction of the particle's wavefunction, a jump, caused by the interaction with the reservoir. $C_\alpha$ projects onto non-orthogonal states, thus $C_\alpha C_\beta \neq 0$ for $\alpha \neq \beta$. Therefore, the measurement we defined does not belong to the class of a Projective-Valued Measure (PVM).  This is in accordance with the modern formulation of a measurement process, which extends the concept of measurements to account for real observations, whose results also depend on the characteristics of the measuring apparatus and procedure. For details on this Positive Operator-Valued Measure (POVM) see Ref.~\cite{Davies76}. Projectors are substituted by an arbitrary number of positive operators, the effects $E_i$, whose sum gives identity $\sum_i E_i = I$ \cite{Kraus83,Holevo01,Wiseman10,Jacobs06}. In the case studied here, the effects are related to a jump within a time interval $dt$ caused by $E_\alpha = dt C^\dagger_\alpha C_\alpha$, or alternatively a no jump event, $E_0=1-\sum_\alpha E_\alpha$. To assure that all effect operators are positive, the time step $dt$ must be sufficiently small. We take care of this fact.  \\

Instead of solving the GKSL equation, in the following we use one of its possible unravellings, the Quantum Monte Carlo Wave Function method \cite{Dalibard92,Molmer93}. The idea of the approach is to generate an ensemble of individual trajectories. Each one can be viewed as a single, possible realization of the dynamics of the wavefunction. Averaging over many such trajectories yields the time dependence of the density operator in accordance with the GKSL equation: $\rho(t) = \overline{|\psi \rangle \langle \psi |}$. The WFQMC method simulates stochastic evolution in which for each time step the quantity $|\phi'(t+\delta t)\rangle$ is calculated, by evolving the state for an infinitesimal time $\delta t$ with the non-unitary Hamiltonian:

\begin{equation}
    \label{eq:hamiltonian}
    H=H_0-\frac{i}{2}\sum_\alpha C^\dagger_\alpha C_\alpha.
\end{equation}

One of two possibilities is then selected: a jump or no-jump event. A jump to the state $|\alpha\rangle$ is selected with the probability:
\begin{equation}
        \label{eq:jump_prob}
        \delta p_\alpha = \delta t \langle \phi' (t) | C^\dagger_\alpha C_\alpha | \phi' (t) \rangle = \gamma \delta t |\langle \alpha | \phi' (t) \rangle|^2.
\end{equation}
 The time-step $\delta t$ has to be sufficiently small to assure that $\sum_\alpha \delta p_\alpha$ is smaller then one. 
 If the jump takes place, the particle's wavefunction changes discontinuously: 
\begin{equation}
     \label{jump}
        |\phi (t+ \delta t) \rangle = |\alpha\rangle.
\end{equation}
The probability of no jump is equal to: 
\begin{equation}
P_0 = 1 -\sum_\alpha \delta p_\alpha. 
\end{equation}

If the `no jump' event takes place, the state is essentially replaced with $|\phi'(t+\delta t)\rangle$. However, since the Hamiltonian (\ref{eq:hamiltonian}) does not preserve the norm, it is first normalized:

    \begin{equation}
    \label{nojump}
        |\phi (t+ \delta t) \rangle = \frac{\left(1-i H \delta t \right) |\phi (t) \rangle}{||\left(1-i H \delta t \right) |\phi (t) \rangle||},
    \end{equation}
The evolution of the wavepacket corresponds thus to a random sequence of jump and no-jump events. In our approach, every jump is interpreted as an act of  measurement. Projection onto states associated with detectors is reminiscent of the reduction of a wavepacket. The nonunitary evolution accounts for the Hamiltonian dynamics of the particle as well as for interaction with the detectors. The hermitian part $H_0$ is the sum of kinetic and potential energy $H_0=\frac{p^2}{2m}+V(x)$. Interaction with the detectors is represented by the nonhermitian term, $\frac{i}{2}  C^\dagger_\alpha C_\alpha$. This term causes a  a kind of `accummulation` of the wavefunction around the detector positions in phase space \cite{Gampel23}.  
In each timestep $\delta t$ every detector contributes to the particle wavefunction, $\phi$, by an amount proportional to $\propto -\frac{1}{2}\gamma \delta t\langle x|\alpha\rangle\langle\alpha|\phi\rangle$. \\
   
Our choice of jump operators $C_\alpha$ fulfills a number of basic assumptions about a sensible detector of position and momentum. First of all, a meter of position should click if the probability to find a particle in its neighbourhood is large. In our case this probability is proportional to the squared overlap of the wavefunction  with the state associated with the detector. Once the meter `fires', the particle wavefunction should be reduced according to the information gained, so the post-measured state is localized around the position of the detector. Choosing the detector states to be Gaussians stands to reason. The  width $\sigma_x$ gives the precision of the measurement. \\

We would like our detectors to be `gentle' to the objects under measurement. By this we do not mean a weak measurement, but we want the particle velocity, assumed to be proportional to the probability density current, to be not significantly affected due to detection. The post-measurement state of the particle should preserve some information about its pre-measurement momentum at the detection point. To this end we equip the detectors at every spatial location with a variety of kinetic momenta by assigning to every Gaussian spatial profile plane waves of momenta $\hbar k_n$. The momenta can take various values as discussed above. The probability of clicking is thus maximal, if both the position \emph{and} momentum of the particle fits one of the detector states. This discussion is obvious if one considers the detector's wavefunction not in  position but in momentum space. The Fourier transform  of a detector state (Eq.(\ref{position})) is:
\begin{equation}
\label{momentum}
    \langle k|\alpha_{mn}\rangle = \left(\frac{2 \sigma^2}{\pi}\right)^\frac{1}{4} e^{-\sigma^2 (k-k_n)^2 + i x_m(k- k_n)},
\end{equation}
a Gaussian superposition of plane-waves of momenta centered around $k_n$. The detector is very sensitive to wavefunctions whose local velocity at $x_m$ is close to $k_n$.\\

\section{Statistical characterization of particle's trajectories}

\begin{figure}[b]
\includegraphics[width=\linewidth]{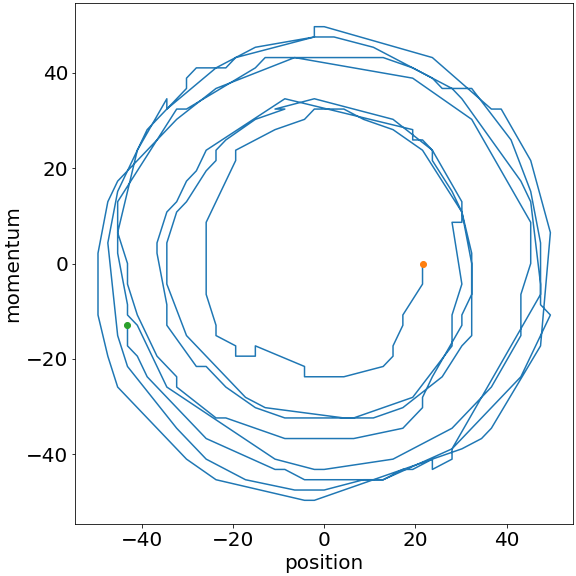}
\caption{\label{fig:trajectory} Sample trajectory in phase space. The start point is marked by an orange point while the end of the simulation is visualized as a green point. A typical trajectory in our setting is always a circular motion with growing radius. Only isolated phase-space positions of a particle are available to the observer. The line is drawn to guide the eye. Detector spacing is $d=2.16$ and $\gamma=1$.}
\end{figure}

\begin{figure}[b]
\includegraphics[width=\linewidth]{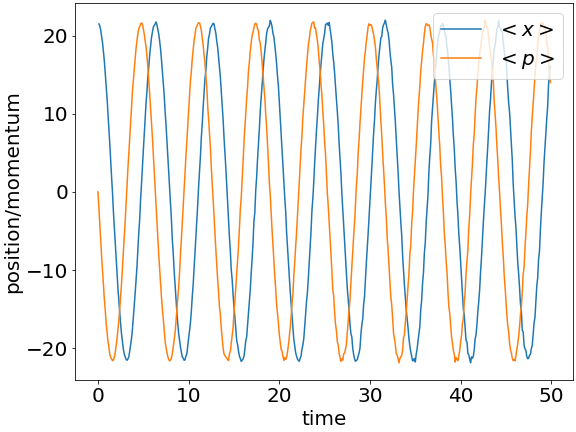}
\caption{\label{fig:avg_pos_mom} Average position and momentum of a particle in the harmonic potential as a function of time. The dependence is equivalent to the classical solution of the harmonic oscillator, giving an sinusoidal motion with the respective frequency and a phase shift of $\pi/2$ between position and momentum. Detector spacing is $d=2.16$ and $\gamma=1$.}
\end{figure}

\begin{figure}[b]
\includegraphics[width=\linewidth]{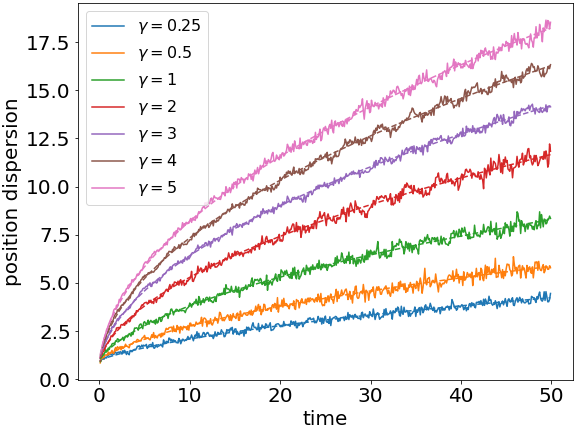}
\caption{\label{fig:std_pos} Dispersion in position (and equivalently, momentum), $\sqrt{\delta_0^2(t)}$ as a function of time. The colours correspond to different values of $\gamma$ for $d=2.16$ The dispersion grows faster for larger $\gamma$. The figure also shows fits of Eq.~(\ref{eq:dissipation}) as dashed lines in corresponding color, which are barely visible because of high agreement.}
\end{figure}

\begin{figure}
    \includegraphics[width=\linewidth]{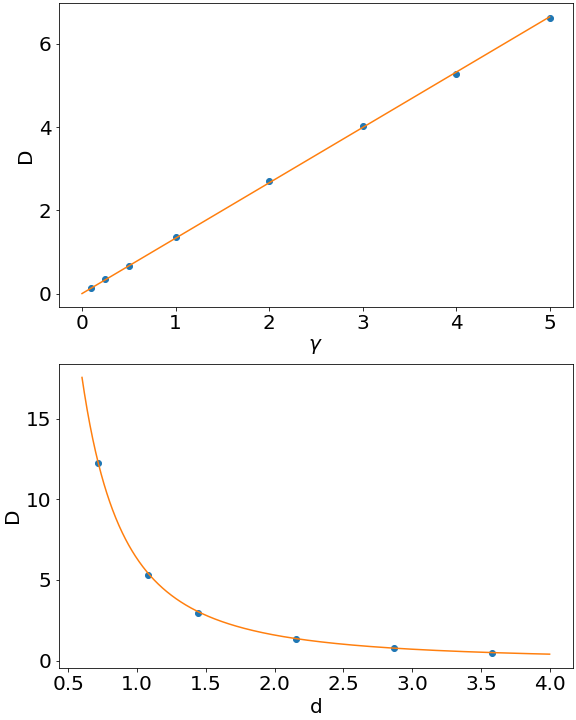}
    \caption{\label{fig:diffusion_coeff} Dependence of the diffusion coefficient $D$ on parameters of detection. The upper panel: dependence of $D$ on $\gamma$ for $d=2.16$. Blue points correspond to simulations while the orange curve is a fit of the linear function, $D \propto \gamma$. In lower panel: dependence of $D$ on  distance $d$ between detectors  for a fixed value of $\gamma =1$. Blue points are the results of simulation, while orange line is the fit of $D \approx (\frac{5}{2})^2 \times \frac{1}{d^2}$ function.}
\end{figure}

\begin{figure}[b]
\includegraphics[width=\linewidth]{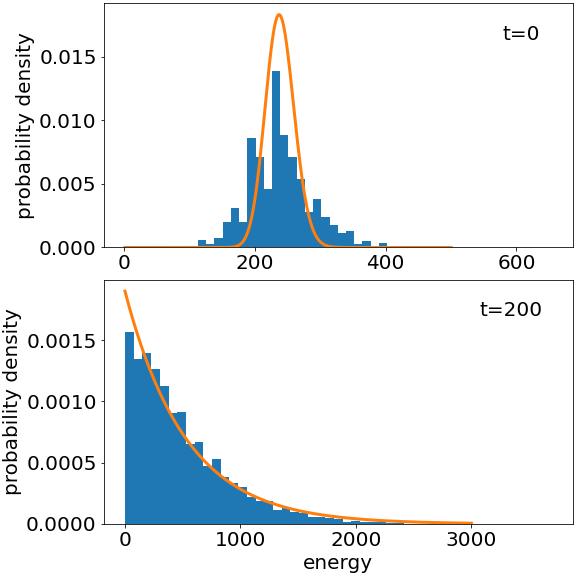}
\caption{\label{fig:energy_dist} Energy distribution for two different times. The upper panel shows a distribution shortly after beginning of the simulation, and the lower panel the distribution at a `late time (after many oscillations). The blue histograms correspond to numerical data while the orange curves are the function Eq.~(\ref{eq:energy_dist})  (top) and an exponential (bottom) fit. Note that the labels $t=0$ and $t=200$ are approximate in the sense that in order to gather sufficient numerical data, for the histogram we consider clicks from a time interval corresponding to one full oscillation.}
\end{figure}

In our work we study `trajectories' of a particle, resulting from the detection process, i.e. sequences of position and momentum measurements of the particle in a harmonic potential $V=\frac{1}{2} m \omega^2 x^2$. We use harmonic oscillator units, i.e as unit of length $a_{ho}=\sqrt{\hbar/m\omega}$, unit of momentum $q_0=\hbar/a_{ho}$, time $\tau_0=1/\omega$, and energy $\varepsilon_0=\hbar \omega$. From now on all quantities are expressed in these units. The hermitian Hamiltonian has the form:
\begin{equation}
\label{eq:unitary_H}
    H_0 = \frac{1}{2} x^2 + \frac{1}{2} p^2,
\end{equation}
placing position and momentum on equal footing.  
We assume that detectors are characterized by a spatial width $\sigma=\sqrt{1/2}$. Selecting this value ensures that the detectors formally have the same width in momentum space. Moreover, as the detectors project into coherent states, the post-measurement uncertainty in position and momentum is minimal according to Heisenberg's principle. Similarly to the hermitian part of the Hamiltonian, the coupling to detectors is symmetric, and position and momentum are on an equal footing. We choose the same numerical value for detector spacing $d_x=d_p=d$. \\

In our calculations, we impose the initial wavefunction of the particle to be identical to one of the Gaussian detector states Eq.(\ref{position}), centered at $(x_0=nd, p_0=0)$. Here $n$ is a natural number, chosen such that $x_0$ is close to 20, so that for different grid densities $d$ we get comparable initial conditions. The particle thus starts with zero velocity at some distance from the potential's minimum. This distance defines the classical amplitude of the harmonic oscillation, and comprises several other detectors ($n \gg 1$), so that the subsequent motion can be monitored with sufficient resolution. In our numerical experiment, we simulate a large number of trajectories, where by trajectory we mean a time series of detection events,  `clicks' of meters at phase-space locations $(x_i,p_i)$ at instants $t_i$. An example of a single realization of the measurement experiment is shown in Fig.~\ref{fig:trajectory}. The particle follows a circular orbit in phase-space, as would be expected for a classical particle. Some random departures from this orbit are clearly visible. Moreover, the radius of the orbit grows slowly in time, i.e. the energy of the observed particle increases. \\

Individual trajectories, resulting from the stochastic process, differ one from another. Their statistical properties are the main objects of our interest. First, we analyze the average phase space trajectory $(\langle x(t)\rangle,\langle p(t) \rangle)$. Using the WFQMC formalism we generate 5000 trajectories for each choice of parameters. The detection events are random and discrete points in time, so to get a mean trajectory we introduce coarse-grained time by dividing the timeline into small intervals, $[t, t+\delta t]$, where $\delta t= 0.1$, and calculate the mean position and momentum for all clicks from the ensemble falling into the interval. Mean trajectories, both in position and  momentum space, show that \emph{on average}, the particle follows a classical path (cf. Fig.~\ref{fig:avg_pos_mom}). Position as well as momentum  oscillate with the harmonic oscillator frequency, and are phase-shifted by $\pi/2$.\\

Deviations of a single realization from the average trajectory are characterized by the second moment of the click distribution, i.e. the dispersion $\delta^2 x(t) =\langle x(t)^2 \rangle -\langle x(t)\rangle^2 $ and  $\delta^2 p(t) = \langle p(t)^2\rangle -\langle p(t)\rangle^2$. Because of the symmetry of the Hamiltonian Eq.(\ref{eq:hamiltonian}) and Eq.(\ref{eq:unitary_H}), the dispersion in position and momentum should be equivalent $\delta^2x=\delta^2p$. Simulations essentially confirm these expectations, which is why in Figure \ref{fig:std_pos}, we only plot the dispersion function $\delta (t) \equiv \sqrt{\delta^2 x}$. The dispersion functions of position and momentum actually differ by a small modulation due to the $\pi/2$ phase shift of position and momentum of the particle. This will be discussed later on in this section.\\

The time dependence, fitted to the numerical results,  is found to be
\begin{equation}
    \label{eq:dissipation}
    \delta^2 (t) \approx D t+\delta_0^2,
\end{equation}
where $D$ is a diffusion coefficient and $\delta_0^2$ the initial dispersion, independent of $\gamma$. $\delta_0^2$ is a result of the initial wavepacket having finite width even when identical to a detector wavefunction. In other words, because of lack of orthogonality, immediately after localization at a detector at position $(x_0,p_0)=(j_0d,k_0d)$, the particle may be captured by a different detector $(x,p)=(jd,kd)$. In our model the probability distribution of a subsequent click of a detector $\alpha_{j,k}$, under the condition that such a click occurs within a short time from the first one, is approximately equal to the discretized Husimi function ${\cal{Q}}(j,k)$ of the initial state \cite{Gampel23}:
\begin{equation}
{\cal{Q}}(j,k) =| \langle \alpha_{j,k}| \alpha_{j_0,k_0}\rangle|^2 =  \frac{ e^{-d^2\left((j-j_0)^2+(k-k_0)^2\right)/2}}{\sum_{j,k}| \langle \alpha_{j,k}| \alpha_{j_0,k_0}\rangle|^2 }.
\end{equation}
According to the discussion above, the  dispersion squared of the initial spatial position of the monitored particle is $\delta_0^2=\sum_{j,k} {\cal{Q}}(j,k) (jd)^2$.
If $d \ll 1$, summation can be substituted by integration, which yields $\delta^2_0 \approx 1$. In the case of numerical results shown in Fig.(\ref{fig:std_pos}), this condition is not satisfied ($d=2.16$), however we surprisingly find that this continuous approximation still works quite well. \\

For large times, the initial dispersion can be neglected, and  Eq. (\ref{eq:dissipation}) indicates that on the top of the harmonic oscillation the particle undergoes Brownian motion. Deviations from the mean trajectory grow as the square root of time, suggesting a diffusion process characterized by the coefficient $D$. Moreover from dimensional analysis it seems that dynamical quantities such as $\delta^2(t)$ should depend on the dimensionless parameter $\gamma t$. Indeed, detailed studies confirm this prediction (see Fig. \ref{fig:diffusion_coeff}, upper panel). This implies that the diffusion coefficient $D$ grows linearly with $\gamma$, which is plausible since this implies more frequent detection of the particle. Similarly, the denser the detector grid, the more detectors there are monitoring the particle, which again leads to a higher detection frequency and larger perturbations of the classical trajectory. In the lower panel of Fig. \ref{fig:diffusion_coeff} we show the dependence of the diffusion coefficient on the detector spacing $d$ for a fixed value of $\gamma=1.0$. Results clearly show that $D$ is inversely proportional to the squared detector spacing. Our numerical experiment allows to postulate the following dependence of the diffusion coefficient on the parameters of the observation process:
\begin{equation}
\label{eq:diffusion}
D \approx 2\pi  \frac{\gamma}{d^2}.
\end{equation}
The analytical formula Eq.(\ref{eq:dissipation}) shows very good agreement with numerical calculations. This formula may also be confirmed by approximate analytical considerations. The diffusion coefficient is related to the squared mean displacement of the walking particle per unit of time, i.e.:
\begin{equation}
D=\gamma \sum_{j,k}e^{-d^2(j^2+k^2)/2}(dj)^2.
\end{equation} 
Using the continuum approximation, $jd=x$, $kd=p$, and $\sum_{j,k} \rightarrow \frac{1}{d^2}\int dx dp$, the diffusion coefficient is equal to:
\begin{equation}
    \label{diffusion}
    D \approx \frac{\gamma}{d^2} \int dx dp \, x^2 e^{-(x^2+p^2)/2} = 2\pi \frac{\gamma}{d^2}.
\end{equation}
We thus recovered Eq. (\ref{eq:diffusion}), which was obtained by fitting to numerical data. \\

A more careful analysis indicates that in addition to the Brownian diffusion characterized by linear growth of the dispersion $\delta^2(t)$, there are small-amplitude oscillations with frequency $2 \omega$. These oscillations can be explained assuming small dephasings of individual trajectories $x = x_0 \cos(t+\delta \varphi)$ with respect to the  average $x = x_0 \cos(t)$. The dephasing gives an oscillatory contribution to the dispersion $\langle x^2 \rangle - \langle x \rangle^2 \approx \delta\varphi^2 \sin^2t$. A similar oscillatory character of dispersion of position and momentum was observed in \cite{Scott01}, where phase space dynamics of a continuously monitored particle in an anharmonic potential is studied. In this work however the dispersion is bounded, contrary to the present result. This is because the authors of \cite{Scott01} study the limit of very frequent and very weak measurements, whereas the present work treats a series of strong measurements at discrete points in time. Each measurement is performed at the `Heisenberg limit', i.e. it minimizes the uncertainty relation:
\begin{equation}
\label{eq:uncertainty}
    \sigma_x \sigma_p = \frac{1}{2}
\end{equation}
Such measurement necessarily introduce growing fluctuations.  Our studies indicate that dispersion of trajectories is model/system sensitive. This fact was also noticed by us in \cite{Gampel23}, where different types of diffusion were found for alternative POVM's of measurement operators.\\

Fluctuations of position and momentum of the particle lead to increasing its energy. It is because of this that when we observe a sample trajectory in phase space, it tends to be a circular motion spiralling outwards (see Fig. \ref{fig:trajectory}). The radius of the circle in phase space increases with time, $r(t)= \sqrt{2\langle E(t) \rangle }$.  It follows directly from Eq.~(\ref{eq:dissipation}) that the average energy of the particle, $\langle E \rangle = \frac{1}{2} \langle x^2 \rangle + \frac{1}{2} \langle p^2 \rangle$,  grows linearly with time:
\begin{equation}
    \label{eq:energy}
    \langle E(t) \rangle =  \delta^2 (t) + E_0   = Dt+(\delta_0^2+E_0),
\end{equation}
where $E_0=\frac{1}{2}(x_0^2+p_0^2)$ is the initial energy of a classical particle at initial position $x_0$ with initial momentum $p_0$. \\

By dividing the energy scale into small intervals $\Delta E$ we can obtain the energy distribution $p_E(t)$ of the ensemble of trajectories as a function of time. This distribution around $t=0$, as obtained from our simulations, is shown in Fig.\ref{fig:energy_dist}. It is a relatively narrow function centered around $E_0$.  Again, as in the case of position dispersion, the initial distribution of energy can be approximately obtained from analytic calculations.
As previously we use the continuum approximation: $j(j_0)d \to x(x_0), k(k_0)d \to p(p_0)$, and ${\cal Q}_{i,j} \to {\cal P}(x,p) = 1/(2 \pi) e^{-(x-x_0)^2/2}e^{-(p-p_0)^2/2}$. If initially the particle is placed at phase-space location $(x_0,p_0)$ then the initial energy distribution is:
\begin{equation}
\label{e_distribution}
    p_E = \int  dx dp \, {\cal P}(x,p) \, \delta\left(E-\frac{1}{2}(x^2+p^2)\right).
\end{equation}
Using that $2 E_0=x^2_0 + p^2_0$  we get:
\begin{equation}
\label{eq:energy_dist}
p_E=e^{-(E+E_0)}I_0(\sqrt{2E_0}\sqrt{2E}),
\end{equation}
where $I_0(z)$ is the modified Bessel function of the first kind. \\

The energy distribution as given by  Eq.(\ref{eq:energy_dist}) is plotted in the upper panel of Fig. \ref{fig:energy_dist}.  Again, the continuous approximation works quite well even for parameters  which do not fully legitimate usage of the formula. We stress that to get the energy histogram we accumulated data from the time interval $0<t<2\pi$, so strictly speaking the histogram does not give the energy distribution exactly at $t=0$, but a distribution averaged over the first period of the oscillation. For large times $t$, this initial energy distribution  evolves into a thermal one:
\begin{equation}
    p_E(t) = \frac{1}{\epsilon(t)} e^{-E/\epsilon(t)},
\end{equation}
The width and mean of this distribution $\epsilon(t)$ depends on time. Setting $\epsilon = kT$ allows us to formally define a temperature for the system, identifying the repeated measurement process with a type of `heating'. The distribution $p_E$ in the thermal regime is shown in the lower panel of Fig. \ref{fig:energy_dist}. The temperature of the ensemble grows with time, and for large times it becomes $kT(t)=\langle E \rangle = \delta^2(t) \approx Dt$. This analytical prediction again agrees well with numerical results.\\

In summary, we studied a quantum particle in an external harmonic potential which is repeatedly monitored by an array of detectors regularly distributed in phase-space. We employed an open system formalism, treating detectors as an external reservoir. Coupling of the particle to the meters is given by jump operators whose action is to project the particle's wavefunction onto coherent states characterizing the detectors. We use the Wave Function Quantum Monte-Carlo method to generate ensembles of time-dependent wavefunctions. We interpret every generated wavefunction as a single realization of the particle's dynamics, which in addition to continuous evolution experiences quantum jumps related to observations.  We show that on average the trajectories follow a classical path. This results is similiar to the one in \cite{Scott01}, where a von Neumann type of coupling between the system -- a nonlinear oscillator -- and meters was considered. The random quantum jumps in position and momentum space introduce fluctuations on top of the harmonic motion. We have shown that these fluctuation have the character of Brownian motion, a diffusive process with dispersion of position and momentum growing linearly with time.  We numerically found the diffusion coefficient and its dependence on the detector clicking rate $\gamma$ and detector spacing $d$. Going back to dimensional units, we see that the diffusion coefficient $D_x$ in position space is proportional to the Planck constant:
\begin{equation}
D_x=4 \pi \gamma \frac{\hbar}{d_x d_p}\sigma^2,
\end{equation}
signifying the quantum character of this process. Again, this is due to the fact that our measurements are performed at the limit set by the Heisenberg uncertainty limit (cf. Eq. (\ref{eq:uncertainty})). $d_x d_p$ is the action equal to the area of an elementary cell in phase space, determined by the detector spacing. \\

Finally, we found that the repeated observation introduces heating of the particle, the energy distribution of the trajectory ensemble at large times becomes thermal and the effective temperature grows linearly in time.\\

Our studies of the system under continuous monitoring and comparison to similar studies \cite{Scott01,Gampel23}, show that observed mean trajectories correspond to the clasical ones, however deviation from the mean, the dispersion, significantly depends on the system studied and details of detection process, in particular on a choice of the Positive Operator-Valued Measure.\\

We do not know whether the particular measurement schemes considered here can ever be realized in practice. However, the model we formulate is fully admissible in view of the present understanding of quantum measurement theory. As such it is legitimate to study its consequences. Paraphrasing the words of prof. Iwo Białynicki \cite{Bialynicki85}: \emph{As to the usefulness of our results, we  have no opinion at all. Perhaps someone else could see whether they are good for anything.}  \\

{\bf Acknowledgements}
The paper is dedicated to Professor Iwo Białynicki-Birula on the occasion of his 90th birthday. Lectures on Quantum Mechanics, given by the Professor at the Physics Department of Warsaw University in the fall semester of 1976, played a very important role in the scientific development of one of us (MG). Moreover, MG  is especially grateful to the Professor for his particular care which MG experienced during his professional life. The Professor and his wife, Zofia, were always eager to offer their friendly help at moments of important decisions. 

We thank Magdalena Załuska-Kotur for illuminating us on the methods of determination of a diffusion coefficient and for pointing to us that $\frac{5}{2} \approx \sqrt{2\pi}$. 

This  work  was  supported  by  the  Polish  National  Science  Centre grant No 2019/32/Z/ST2/00016,  through the project MAQS under QuantERA, which has received funding  from  the  European  Union’s  Horizon  2020  research and innovation program under Grant Agreement No 731473.\\

\end{document}